\documentclass[aps,10pt,floatfix,twocolumn,showpacs]{revtex4-1} 
\usepackage{color,soul}
\usepackage{graphicx}
\usepackage{siunitx}
\usepackage[breaklinks=true,colorlinks=true,linkcolor=blue,urlcolor=blue,citecolor=blue]{hyperref}

\usepackage{amsmath, amsthm, amssymb}
\usepackage{amsthm}
\usepackage{multirow}
\usepackage[normalem]{ulem}
\usepackage{soul}
\usepackage{makecell}
\usepackage[table]{xcolor}
\definecolor{Gray}{gray}{0.85}
\definecolor{LightCyan}{rgb}{0.88, 1, 1}
\definecolor{Apricot}{rgb}{0.98, 0.81, 0.69}
\usepackage{threeparttable}
\newcommand{\be}{\begin{equation}}
\newcommand{\ee}{\end{equation}}
\newcommand{\bea}{\begin{eqnarray}}
\newcommand{\eea}{\end{eqnarray}}

\usepackage[rightcaption,]{sidecap}
\sidecaptionvpos{figure}{c}

\begin{document}

\title{
Glassy dynamics in two-dimensional ring polymers: size versus stiffness polydispersity}
\author{Rahul Nayak}
\email{rahulnayak@imsc.res.in}
\affiliation{The Institute of Mathematical Sciences, C.I.T. Campus,
Taramani, Chennai 600113, India}
\affiliation{Homi Bhabha National Institute, Training School Complex, Anushakti Nagar, Mumbai, 400094, India}
\author{Pinaki Chaudhuri}
\email{pinakic@imsc.res.in}
\affiliation{The Institute of Mathematical Sciences, C.I.T. Campus,
Taramani, Chennai 600113, India}
\affiliation{Homi Bhabha National Institute, Training School Complex, Anushakti Nagar, Mumbai, 400094, India}
\author{Satyavani Vemparala}
\email{vani@imsc.res.in}
\affiliation{The Institute of Mathematical Sciences, C.I.T. Campus,
Taramani, Chennai 600113, India}
\affiliation{Homi Bhabha National Institute, Training School Complex, Anushakti Nagar, Mumbai, 400094, India}
\date{\today}
\begin{abstract}
Soft glassy materials often consist of deformable objects. Here, we use a two-dimensional assembly of semi-flexible ring polymers as a model system to investigate how polydispersity in particle stiffness or size influences the onset of glassy dynamics. In simulations at fixed polydispersity ($30\%$), we find that stiffness dispersity drives most rings into elongated conformations at high densities, leading to orientationally ordered structures that cause dynamical slowing down. In contrast, size dispersity generates a bimodal population: small rings remain circular and act as rigid inclusions, while large rings elongate, producing frustration that delays arrest. Real-space maps of bond relaxation reveal strikingly different pathways of dynamical heterogeneity, with long-lived domains persisting under stiffness dispersity but rapidly percolating relaxation under size dispersity. Moreover, local correlations between ring shape, orientational order, and mobility show that stiffness dispersity produces dynamics that are strongly structure-sensitive, whereas size dispersity activates motion from both circular and elongated populations. By linking microscopic deformability to emergent glassy dynamics, this study identifies how the nature of polydispersity controls the relaxation pathways of soft glasses. 
\end{abstract}

\maketitle

\section{Introduction}\label{SEC-1}

Soft glassy materials~\cite{cipelletti2005slow, bandyopadhyay2006slow, binder2011glassy} are abundant in nature, in the biological world and also used in many applications and many of these materials have deformable objects as constituents. To date, most theoretical models of soft glasses have treated the constituent particles as point-like, interacting as hard or soft spheres. Modelling deformable particles is challenging~\cite{likos2006soft, manning2023essay}, and simulations of large assemblies of such deformable objects are often computationally demanding. Nevertheless, recent studies have begun to explore how particle deformability influences the formation of disordered amorphous structures~\cite{mukhopadhyay2011packings, makse2000packing, batista2010crystallization, boromand2018jamming, gnan2019microscopic}. Introducing deformability and softness into particles can induce new phases at high packing densities that are unattainable with rigid hard-sphere particles, due to shape distortion and many-body interactions.

In nature as well as in synthetic materials, constituents are typically polydisperse, i.e., there is a distribution of sizes, shapes, charges, or other properties across the particles. This ubiquity of polydispersity has prompted extensive studies on how it influences material properties. It is well established that beyond a certain threshold of polydispersity in size, crystallization of particles is strongly suppressed via structural frustration~\cite{pusey1986phase, mcrae1988freezing, phan1998effects, auer2001suppression, schope2007effect}. Instead of crystallizing, a highly polydisperse system is predicted to form either an equilibrium amorphous state~\cite{chaudhuri2005equilibrium} or to undergo fractionation~\cite{fasolo2003equilibrium}. Thus, systems with sufficiently large polydispersity are typically good glass formers. Indeed, polydisperse colloidal and molecular systems have been extensively used to investigate diverse properties of glassy states~\cite{weeks2000three, weysser2010structural, murarka2003diffusion, zaccarelli2015polydispersity, behera2017effects, heckendorf2017size, laudicina2023competing}. Such systems have even enabled the development of efficient algorithms to produce ultra-stable glasses~\cite{ninarello2017models}. Moreover, increasing the degree of polydispersity tends to delay the onset of glassiness to higher density (or lower temperature) and typically produces less fragile glass formers~\cite{abraham2008energy}. In other words, at any density/temperature, a system with greater polydispersity relaxes faster and exhibits less dynamic heterogeneity than a more monodisperse system~\cite{kawasaki2007correlation, abraham2008energy, kawasaki2011structural}. Consistent with this, recent experiments and simulations have shown that even a jammed colloidal suspension can be re-fluidized by broadening the particle size distribution, highlighting that greater size dispersity facilitates particle rearrangements~\cite{biswas2024influence}. 

In addition to size polydispersity, deformability polydispersity, i.e. variations in the elastic stiffness of particles, has gained attention in the context of glassy systems. For instance, studies of soft colloids and hydrogels have shown that particle deformability can lead to diverse mechanical responses, impacting the onset of glassiness and the structural organization of the system~\cite{zhao2024elasticity, pellet2016glass}. Variations in particle stiffness result in differential deformation under compression: softer particles deform more readily, which allows the system to achieve higher effective packing fractions. This effect has been observed to reduce the extent of crystallization and to enhance the glass-forming ability in suspensions of deformable particles~\cite{he2021deformability, li2022softness}. Systems with a wide spread in particle stiffness (high deformability polydispersity) also exhibit a broad distribution of local stresses, which in turn affects the mechanical stability and dynamical response of the packed structure~\cite{liu2019effects, behera2017effects}. These findings underscore that softness heterogeneity can profoundly influence how and when a dense suspension becomes glassy.

Polymer-based colloidal systems, characterized by their soft and tunable interactions, have been widely used as model frameworks to investigate the behavior of soft colloids. Representative examples include linear polymer chains (flexible or semi-flexible)\cite{louis2000can}, star and dendritic polymers\cite{gotze2004tunable, likos2006soft}, microgels (polymer networks that can swell or shrink)\cite{ninarello2019modeling}, and, more recently, topologically constrained ring polymers\cite{gnan2019microscopic,roy2022effect}. Within this class, ring polymers hold particular significance for elucidating glassy dynamics in deformable systems, because their closed-loop topology suppresses the reptation mechanism that normally governs the relaxation of linear polymers. This restriction of the usual entanglement relaxation mode dramatically influences ring polymer dynamics, often leading to unique arrested states at high densities. For example, concentrated ring polymers can form stacked clusters of rings in quasi-2D confinement~\cite{slimani2014cluster}, and in general they exhibit slow dynamics very different from their linear counterparts. The absence of chain ends and the resulting distinctive relaxation pathways make ring polymers valuable minimal models for exploring dense, biologically relevant structures such as chromatin (which can be viewed as a system of closed loops). The interactions in these soft polymer systems can be tuned at multiple levels: from effective coarse-grained interactions between centers of mass, to local bending/stretching stiffness along the polymer backbone. The stiffness of each polymer strongly affects its packing conformation and the overall material properties of the assembly.

Notably, recent simulation work has demonstrated a re-entrant melting transition in two-dimensional assemblies of soft colloidal particles modeled as ring polymers at very high packing fractions~\cite{gnan2019microscopic}. In this 2D system, increasing the density initially causes a glass transition, but beyond a critical extreme density the system melts back into a fluid-like state due to the rings' deformability. As the disks (ring polymers) become highly compressed, their effective interaction is no longer pairwise-additive; instead a many-body repulsion emerges, and the nature of deformation shifts from localized particle flattening to a more distributed deformation field spanning multiple particles~\cite{vsiber2013many}. This collective deformation at super-high density relieves local stress and is consistent with the onset of a hard-core-like behavior, thus producing the re-entrant fluidization.  In particular, simulations of deformable 2D polygons have revealed pronounced particle shape changes at the jamming transition, and these have been connected to what one might expect in thermal systems approaching the glass transition~\cite{boromand2018jamming, treado2021bridging}. These findings emphasize that when particles can change shape or deform, new pathways for packing and relaxation become available, often altering the classic glass or jamming phenomenology.

In our previous work~\cite{ghosh2024onset}, we investigated the onset of glassy dynamics in 2D ring polymer assemblies that were monodisperse in size, focusing on how ring stiffness and crowding give rise to dynamic slowdown and structural organization. Through coarse-grained molecular dynamics simulations, we demonstrated that flexible rings undergo glass formation primarily via crumpling‑induced crowding, forming dense, globular structures without large-scale orientational order, whereas stiffer rings maintain polygonal conformations and develop locally ordered domains at high densities. We showed that, in 2D, glassy arrest occurs via deformability and packing constraints, not threading as in 3D, and that stiffer rings reach glassiness at lower densities~\cite{ghosh2024onset}. In summary, our earlier study established that ring polymer stiffness is a key control parameter for glass formation in 2D, and that the route to arrest for flexible vs. stiff rings is qualitatively different.

Building on these findings, the current study aims to understand how the introduction of disorder in the form of polydispersity affects the dynamical behavior of 2D ring polymer assemblies. Polydispersity can be introduced in two distinct ways in this system: (i) by having a distribution of ring flexibilities (bending stiffnesses), or (ii) by having a distribution of ring sizes (contour lengths). We investigate both types of polydisperse constructions and perform a comparative analysis of their structural and dynamical properties. By fixing the polydispersity to the same overall level in each case, we can directly contrast how flexibility polydispersity versus size polydispersity influences the onset of glassy dynamics. Through this comparative study, we seek to elucidate the mechanisms by which either form of polydispersity drives  glass formation in deformable 2D systems. Our work thus provides new insights into the complex behavior of two-dimensional soft materials and suggests design principles for materials where one can exploit polydispersity (in size or stiffness) to achieve controlled kinetic stability and specific glassy properties.

\begin{figure*}
	\centering
	\includegraphics[width=1.6\columnwidth]{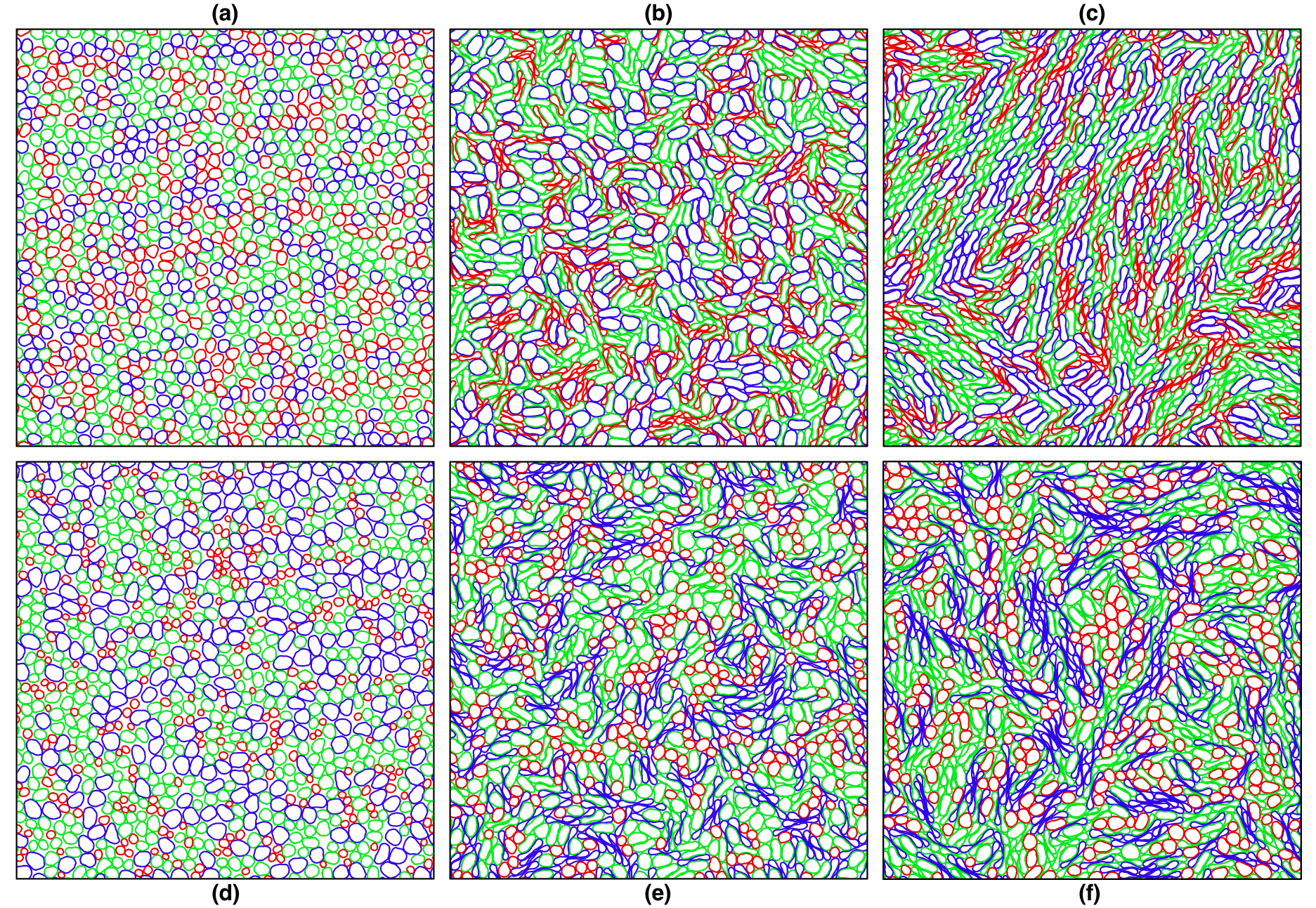} 
	\caption{Snapshots at three different densities, viz. $\rho=0.085, 0.168, 0.226$ (from left to right): (a)-(c) stiffness polydisperse: red, green, blue mark small,  medium and large stiffness. (d)-(f)  size polydisperse: small, medium-sized and large rings are respectively marked in red, green, and blue.}
	\label{snapshots}
\end{figure*}

\section{Modeling and Methods}\label{SEC-II}

\subsection{Model}

We consider the ring polymer to be consisting of $n_m$ monomers, and these monomers interact
via  a combination of Weeks-Chandler-Andersen (WCA) potential
\begin{equation}
    U_{LJ}(r)=
    \begin{cases}
    4\epsilon[(\frac{\sigma}{r})^{12}-(\frac{\sigma}{r})^{6}]+\epsilon & \text{if }r\le 2^{\frac{1}{6}}\sigma\\
    0 & \text{if }r>2^{\frac{1}{6}}\sigma
    \end{cases}
\end{equation}
and finite extensible nonlinear elastic (FENE) potential
\begin{equation}
    U_{FENE}(r)=-\epsilon k_{F}R_{F}^{2}\ln [1-(\frac{r}{R_{F}\sigma})^{2}] \quad \text{if }r<R_{F}\sigma
\end{equation}
between bonded monomers, where $\sigma$ is the diameter of each monomer (and the unit of length), $\epsilon$ is the unit of energy, $k_{F}=15$ is the spring constant, and $R_{F}=1.5$ is the maximum extension of the bond ~\cite{smrek2020active}. To model the flexibility of the ring, we have used an angular potential
\begin{equation}
    U(\theta)=K_{\theta}(1-cos(\theta-\pi))
\end{equation}
where $K_{\theta}$ is the angular stiffness.

In this study, we consider two different kinds of systems -- (i) one which we call stiffness polydisperse, i.e. different rings have different angular stiffness $K_\theta$, with a mean $\overline{K_\theta}=100$, and (ii)  the other is size polydisperse, i.e. the rings have different number of monomers, with a mean size $\overline{n_m}=100$. For the stiffness polydisperse system, the rings have $n_m=100$ monomers, and for the size polydisperse system, all the rings have $K_\theta=100$.  For the stiffness poyldisperse system,  $K_\theta$ for the rings are sampled from a uniform distribution. Similarly, in the size polydisperse system,  the ring sizes are sampled from an uniform distribution for $n_m$. Polydispersity is defined as the ratio of the standard-deviation of the distribution to its mean, and represented in percentage form. In this study, we consider $30\%$ polydispersity for both stiffness as well as the size polydisperse systems.

\subsection{Methods}

The system of ring polymers ($N=1000$ rings) is initialized at dilute density by randomly placing non-overlapping polymer rings within the simulation box and the equilibriated at $T = 3.0$ and at fixed volume, over a time window of $10^5$.  NPT simulations are subsequently done at $T = 1.0$ and at several target pressures for an additional time window of $10^5$ to obtain denser thermal assembles, using the Nosé-Hoover barostat to maintain the desired pressure. The eventual equilibrium structural and dynamical properties are investigated within the NVT ensemble, by fixing the density at the average density corresponding to the target pressure. Prior to the NVT production runs, additional NVT equilibration runs are conducted over a time window of $10^7$. The production run spans a time window of $10^7$. We average over $40$ time origins, using $4$ independent trajectories. In all cases, the numerical integration is done using a timestep of $0.005$. 
All simulations are performed using the LAMMPS molecular dynamics software \cite{plimpton1995fast}, with all quantities reported in reduced Lennard-Jones (LJ) units.

\section{Results}\label{SEC-III}

\subsection{Structural properties}

\subsubsection{Spatial organisation of polydisperse rings}

In Fig.~\ref{snapshots}, we show representative snapshots sampled from equilibrium trajectories of systems with 30\% stiffness polydispersity (subplots (a)–(c)) and 30\% size polydispersity (subplots (d)–(f)) at three different densities, $\rho=0.085,\,0.168,\,0.226$, where $\rho$ is the number of monomers per unit area of the simulation box.  At the lowest density, $\rho=0.085$ (Fig.~\ref{snapshots}(a),(d)), both systems are disordered and spatially homogeneous, with only mild ring deformation. In the stiffness-polydisperse system, softer rings (red) are able to adjust slightly to their neighbors. In the size-polydisperse system, the small rings (red) preferentially occupy interstitial spaces between medium-sized (green) and large (blue) rings. At $\rho=0.168$ (Fig.~\ref{snapshots}(b),(e)), deformation becomes significant. In the stiffness-polydisperse case, many rings elongate, with softer rings (red) showing pronounced shape changes and local alignments begin to emerge. In the size-polydisperse case, the largest rings (blue) deform strongly from circular to elongated shapes, while medium-sized rings (green) show intermediate deformation. Small rings (red), however, remain nearly circular and act as rigid inclusions, thereby disrupting any emergence of local ordering. At the highest density, $\rho=0.226$ (Fig.~\ref{snapshots}(c),(f)), the contrast between the two systems is striking. In the stiffness-polydisperse system, most rings elongate into rod-like conformations and assemble into extended, domain-like structures with local smectic ordering, while stiffer rings provide a structural backbone. In the size-polydisperse system, large rings become highly anisotropic and align locally, but the presence of small, nearly rigid rings prevents long-range order, leaving the system frustrated with only short-range correlations.

\begin{figure*}
	\centering
	\includegraphics[width=1.6\columnwidth]{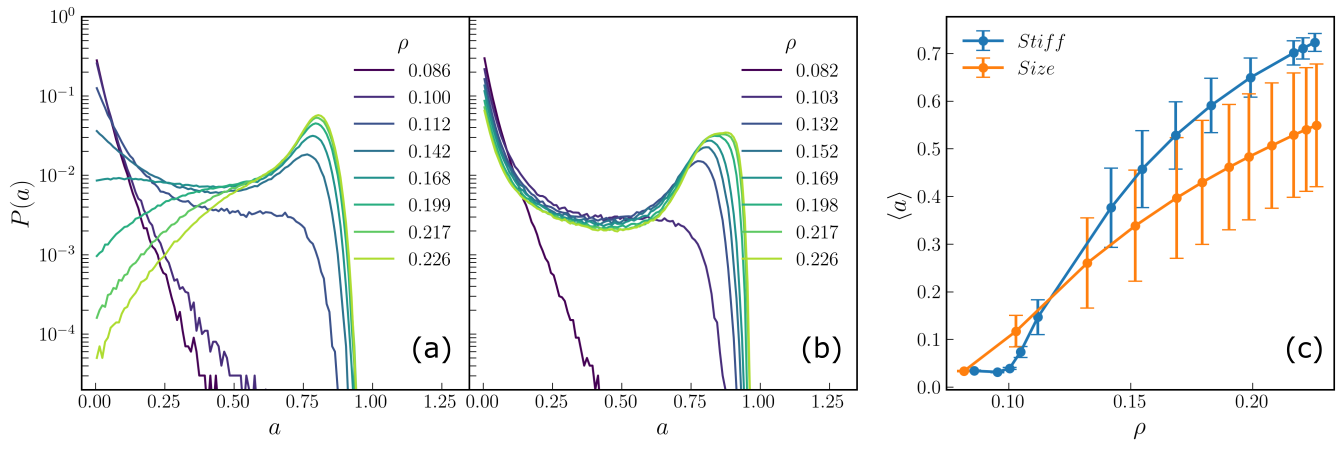}\\
	\caption{(Top). Distribution of asphericity of rings, $P(a)$, measured for different values of density as marked, corresponding to (a) stiffness  and (b)  size polydisperse systems. (c) Variation of the average asphericity, $\langle{a}\rangle$, with density $\rho$ for the two kinds of polydisperse systems.}
	\label{asph}
\end{figure*}

These visual observations are consistent with the density-dependence of the average radius of gyration, $R_g$, shown in Fig.~S1 (See Supplementary Information). With increasing $\rho$, the overall $R_g$ decreases in both systems, reflecting progressive compaction of the rings. However, the decrease is more uniform across all rings in the stiffness-polydisperse system, whereas in the size-polydisperse case, the smallest rings maintain nearly constant $R_g$ across densities. This confirms that in the latter case the small rings act as rigid circular inclusions, while the larger rings undergo substantial deformation. In the following, we quantify these shape changes more systematically using the distributions of asphericity (Fig.~\ref{asph}) and thereafter quantify possible existence of local orientational order.

\begin{figure*}
	\centering
	\includegraphics[width=1.6\columnwidth]{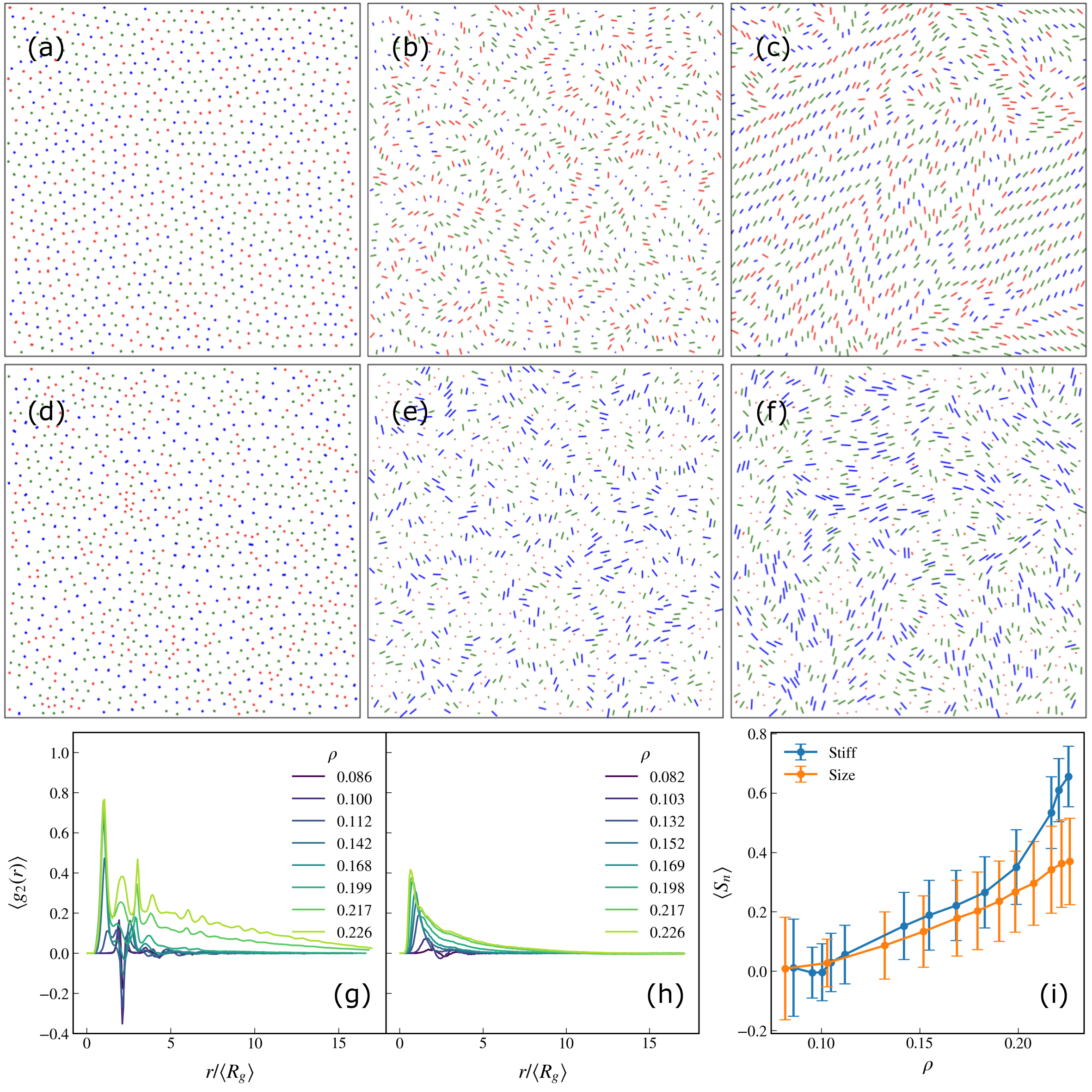}\\
	\caption{Rod representation of the snapshots shown in Fig.\ref{snapshots}, for  (a)-(c) stiffness (d)-(f) size polydisperse systems  at  $\rho=0.085, 0.168, 0.226$ (from left to right); length of each rod corresponds to largest eigen-vector of the gyration tensor multiplied by the asphericity of the ring. (g)-(h) Orientational correlation function, $g_2(r)$, plotted respectively for the stiffness and size polydisperse systems at the listed densities; distances are scaled by the density-dependent average radius of gyration, $\langle{R_g}\rangle$. (i) Variation of average local orientational order parameter $\langle{S_n}\rangle$ with density $\rho$.}
	\label{g2_Sn}
\end{figure*}


\subsubsection{Asphericity Distribution}

The behavior of deformable 2D rings under densification reveals a complex interplay between size, deformation, and effective stiffness. As seen in Fig.~\ref{snapshots}, the ring shapes evolve strongly with increasing density. This can be quantified by the asphericity $a$, computed from the gyration tensor, $G_{mn} = \frac{1}{N}\sum_{i=1}^{N}(r_{m}^{(i)}-R_{m})(r_{n}^{(i)}-R_{n})$, where $r$ and $R$ denote monomer and center-of-mass coordinates and $m$ and $n$ represent Cartesian coordinate indices.  In two dimensions, $G_{mn}$ has eigenvalues $\lambda_{1}^2$ and $\lambda_{2}^2$, from which the asphericity is defined as
\begin{equation}
a= \frac{(\lambda_{1}^{2}-\lambda_{2}^{2})^2}{(\lambda_{1}^{2}+\lambda_{2}^{2})^2}.
\end{equation}

The distributions $P(a)$ shown in Fig.~\ref{asph}(a),(b) provide key insights into deformation. In both size- and stiffness-polydisperse systems, increasing density shifts $P(a)$ toward larger $a$ and broadens the distribution, indicating progressively more elongated rings. However, the details differ strongly. For the stiffness-polydisperse system (Fig.~\ref{asph}(a)), $P(a)$ evolves from a narrow peak at $a\approx 0$ with a small tail at low density to a unimodal distribution centered at large $a$ at high density. Thus, nearly all rings become rod-like under compaction. Correspondingly, the mean asphericity $\langle a \rangle$ (Fig.~\ref{asph}(c)) rises steeply with $\rho$, while the variance decreases, showing that deformation becomes homogeneous, at high density, across the range of $K_\theta$ that constitutes the system. This trend is similar to what was observed for rings which are monodisperse in stiffness~\cite{ghosh2024onset}. In contrast, the size-polydisperse system exhibits distinctly bimodal $P(a)$ at high densities (Fig.~\ref{asph}(b)): small rings remain nearly circular ($a \sim 0$) while large rings become highly elongated, producing a second peak at large $a$. Medium-sized rings populate the intermediate regime. As a result, $\langle a \rangle$ increases with $\rho$ more slowly than in the stiffness-polydisperse case, and the variance remains large at high $\rho$ (Fig.~\ref{asph}(c)). This confirms that shape diversity persists due to the rigid small rings acting as inclusions.

\subsubsection{Spatial correlations}
As shown in Fig.~\ref{snapshots}(c), we observe that at high density, once rings deform into rod-like shapes, they begin to align with their neighbors. Similar orientational ordering was previously reported for monodisperse rings~\cite{ghosh2024onset}.  To visualize this systematically, we compute the largest eigenvector of the gyration tensor for each ring and multiply its magnitude with its asphericity value. The resulting rod-representation is shown in Fig.~\ref{g2_Sn}(a)–(c) for stiffness polydispersity and Fig.~\ref{g2_Sn}(d)–(f) for size polydispersity. Dots correspond to nearly circular rings, while rods indicate elongated rings with orientation. The visualization reveals that at high density ($\rho=0.226$), the stiffness-polydisperse system develops domains with smectic-like bundling that extend over large lengthscales [Fig.~\ref{snapshots}(c)], whereas only short-range alignments are visible in the size-polydisperse system [Fig.~\ref{snapshots}(f)]. Even at intermediate density ($\rho=0.168$), extensive linear patches of aligned rings are evident in the stiffness-polydisperse system [Fig.~\ref{g2_Sn}(b)], highlighting the onset of domain formation. To quantify these trends, we compute the orientational correlation function~\cite{Frenkel2000},
\begin{equation}
g_{l}(r) = \langle \cos(l(\theta(0)-\theta(r))) \rangle,
\end{equation}
where $\theta$ is the angle of the largest eigenvector with respect to the $x$-axis. Focusing on $g_2(r)$, which probes nematic correlations~\cite{Frenkel2000, ghosh2024onset}, we find that the stiffness-polydisperse system exhibits correlations extending over increasingly larger distances with density [Fig.~\ref{g2_Sn}(g)], while correlations remain short-ranged in the size-polydisperse system [Fig.~\ref{g2_Sn}(h)].

A complementary measure is the local orientational order parameter~\cite{zheng2014structural},
\begin{equation}
S_n = \langle S_n^\ell \rangle = \left\langle \frac{1}{n_b}\sum_{j=1}^{n_b} \cos(2\Delta\theta_j) \right\rangle,
\end{equation}
where $S_n^\ell$ is defined per ring from its $n_b$ nearest neighbors, and $\Delta\theta_j$ is the relative angle with neighbor $j$. Neighbors are defined as rings sharing one monomer within $3\sigma$. As shown in Fig.~\ref{g2_Sn}(i), $\langle S_n \rangle$ increases with density in both systems, but grows more sharply for stiffness polydispersity, exceeding 0.6 at $\rho>0.2$, whereas in size polydispersity it plateaus at lower values with larger variance arising out of shape heterogeneity. These results highlight the distinct mechanisms underlying order. In size-polydisperse assemblies, geometric frustration from small, rigid rings prevents large rings from achieving coherent alignment, resulting in a spatially disordered assembly. In stiffness-polydisperse systems, by contrast, densification leads to uniform rod-like shaping of most of the rings which inturn promotes aligned odering in the form of extended domains. 

\begin{figure*}[htb]
	\centering
	\includegraphics[width=1.7\columnwidth]{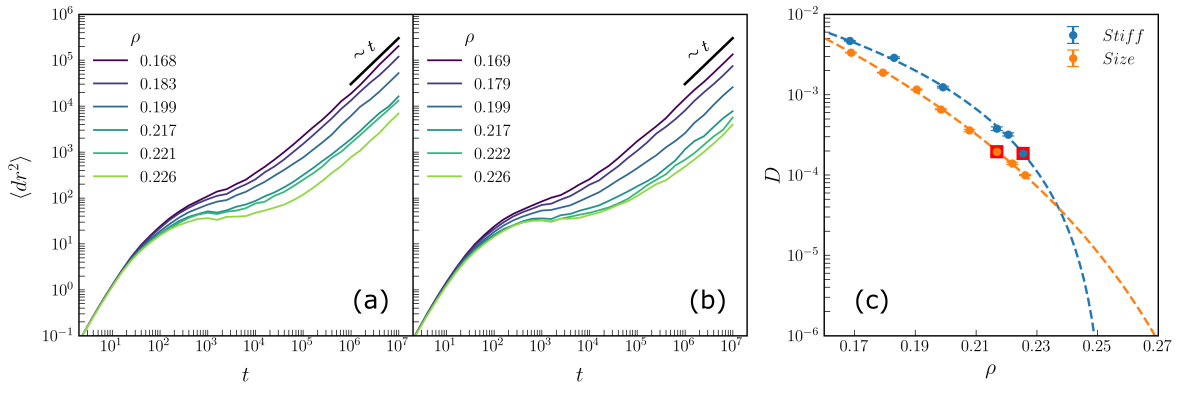}
	\caption{Time evolution of the mean square displacements (MSD) of center of mass of the rings, $\langle{dr^2}\rangle$ for (a) stiffness polydisperse, (b) size polydisperse, across various densities as marked. (c) Corresponding variation of diffusion coefficient ($D$) with density ($\rho$). Dashed lines correspond to fits with $(\rho_g-\rho)^\beta$; see text for estimates of $\rho_g$ and $\beta$ for each case. The red squares mark the iso-$D$ state points explored in Fig.\ref{isoD}.}
	\label{MSDplots}
\end{figure*}

\subsection{Dynamical properties}

\subsubsection{MSD of center of mass of rings}

To characterize the dynamics of the system, we measure the mean-squared displacement (MSD) of the centers of mass (COM) of the rings,
\begin{equation}
dr^{2}(t)=\frac{1}{N}\sum_{i=1}^{N}\left|\Vec{R_{i}}(t)-\Vec{R_{i}}(0)\right|^{2},
\end{equation}
where $N$ is the total number of rings and $\Vec{R}_{i}(0)$ and $\Vec{R}_{i}(t)$ are the coordinates of the center of mass of the $i^{\mathrm{th}}$ ring at times 0 and $t$, respectively. The COM self-diffusion coefficient $D$ is obtained from the ensemble-averaged MSD --
\begin{equation}
D=\lim_{t\to\infty}\frac{\langle dr^{2}(t)\rangle}{4t},
\end{equation}
with $\langle\cdots\rangle$ corresponding to the averaging done over multiple time origins and independent trajectories.

Figure~\ref{MSDplots}(a),(b) shows the MSD data for stiffness- and size-polydisperse systems, respectively, across a range of densities. With increasing $\rho$, dynamical signatures of glassy onset are evident in both cases. Following the initial ballistic regime, $\langle dr^{2}\rangle\sim t^{2}$, a plateau develops at intermediate times and the duration of the plateau grows longer with density, indicating increased caging by neighboring rings. At long times, diffusive behavior $\langle dr^{2}\rangle\sim t$ is recovered for the range of densities reported in  Figure~\ref{MSDplots}(a),(b). For even larger densities, the diffusive regime is not reached within accessible simulation timescales (see Fig.~S2). This absence of long-time diffusion marks the non-ergodic behaviour characteristic of glassy systems.

In Fig.~\ref{MSDplots}(c), the diffusion coefficient $D$ extracted from the MSD (see Fig.~S3) is plotted as a function of $\rho$ for both systems. As expected, $D$ decreases monotonically with density. Fitting the data with $D \sim (\rho_g-\rho)^{\beta}$ gives $\rho_g=0.317, 0.254$ and $\beta=7.2, 2.9$, respectively for the size-polydisperse and the stiffness-polydisperse systems; $\rho_g$ is the estimated density where non-ergodicity would set in. Thus, from these fits, the stiffness-polydisperse system is estimated to undergo dynamical arrest at lower density and exhibits a steeper, more fragile-like slowdown, as compared to the size-polydisperse system.

These dynamical results  coupled with  the structural features identified in Figs.~\ref{snapshots}–\ref{g2_Sn} leads to the conclusion that in stiffness polydispersity, the development of smectic domains promotes cooperative slowing down, {while in size polydispersity, geometric frustration delays the dynamical arrest}. To probe these differences more robustly, we next analyze the bond correlations.

\subsubsection{Bond correlation function}
In two-dimensional (2D) glassy systems, translational degrees of freedom are strongly affected by long-wavelength Mermin–Wagner fluctuations~\cite{vivek2017long}. In contrast, bond correlations are largely unaffected and thus provide a more robust measure of structural decorrelation in two dimensions~\cite{shiba2016unveiling, illing2017mermin}. We therefore analyze bond correlations as a function of density for the two polydisperse systems. 

\begin{figure}
	\centering
	\includegraphics[width=0.9\columnwidth]{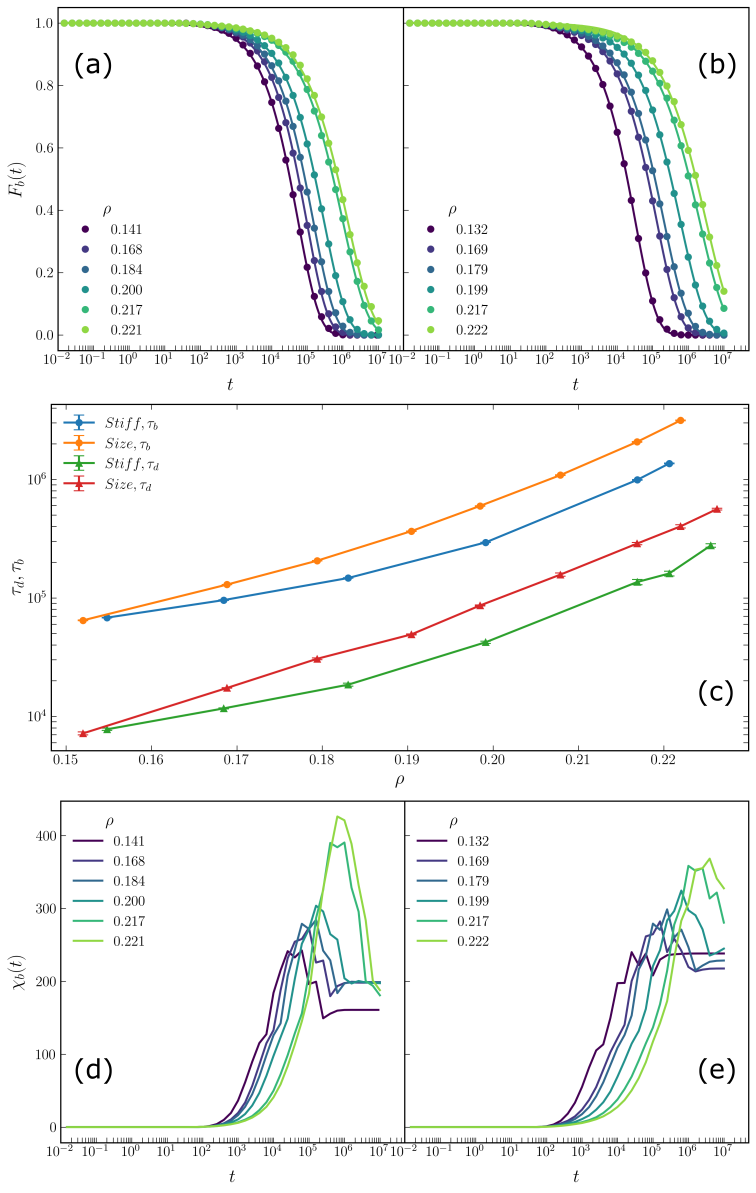}
	\caption{Time evolution of bond correlation function, $F_b(t)$, for (a) stiffness and (b) size polydisperse systems; data shown with points and the lines correspond to fits with stretched exponential functions (see main text for details).  (c) Density dependence of relaxation timescale, $\tau_b$, estimated from $F_b(t)$, shown along with the diffusive timescale $\tau_D$ (defined in the main text), for the two systems as marked. (d) Fluctuations in bond correlations, $\chi_b(t)$, as a function of time, corresponding to data shown in (a)-(b) respectively.}
	\label{bondcorr}
\end{figure}

The center of mass of two rings are defined to be {\em virtually bonded} if any monomer from one ring lies within a distance of $3\sigma$ of a monomer of another. At the initial time $t=0$, the number of such bonds is $N_b(0)$. As time evolves, bonds break due to ring rearrangements caused by thermal fluctuations, and the number of surviving bonds is $N_b(t)$. In our analysis, we deem that the virtual bond is broken at any time $t$ if the distance between the COM-s of the bonded rings is stretched beyond $1.5\delta{R}$, where $\delta{R}$ is the initial distance between the two COM-s at $t=0$. This would imply that the two neighbouring rings have now moved beyond typical nearest-neighbour distances (see data for pair correlation functions in Fig.~S4 in Supplementary Information). The bond correlation function is then defined as~\cite{shiba2012relationship, goto2023unraveling}
\begin{equation}
F_b(t) = \left\langle \frac{N_b(t)}{N_b(0)} \right\rangle ,
\end{equation}
where $\langle \cdots \rangle$ denotes averaging over 50 independent initial configurations. $F_b(t)$ thus represents the fraction of surviving neighbors as a function of time. The measured $F_b(t)$ is shown in Fig.~\ref{bondcorr}(a),(b). With increasing density, $F_b(t)$ decays more slowly in both systems. A characteristic relaxation time, $\tau_b$, is extracted by fitting $F_b(t)$ with a stretched exponential,
\begin{equation}
F_b(t) = \exp\!\left[-\left(\frac{t}{\tau_b}\right)^\beta\right],
\end{equation}
where $\beta=1$ corresponds to simple exponential relaxation and $\beta<1$ indicates a distribution of relaxation times, characteristic of heterogeneous glassy dynamics. The resulting $\tau_b$ is plotted in Fig.~\ref{bondcorr}(c). As expected, $\tau_b$ increases strongly with $\rho$. Notably, the stiffness-polydisperse system exhibits a steeper growth, signaling fragile-like behavior, whereas the size-polydisperse system grows more smoothly, consistent with a less fragile glass~\cite{kawasaki2011structural}. Comparison with the diffusive timescale $\tau_D=\langle R_g^2 \rangle/(4D)$~\cite{goto2023unraveling}, also shown in Fig.~\ref{bondcorr}(c), reveals that $\tau_b$ exceeds $\tau_D$ by more than an order of magnitude across all densities, consistent with previous studies~\cite{goto2023unraveling}. However, more importantly, the growth in $\tau_b$, which stems from neighbourhood changes and thus a collective process, is very similar to thr growth in $\tau_D$, which captures single-ring translational motion. Thus the bond-decorrelation measurements confirm the difference in the nature of dynamical slowing down across the two polydisperse systems.

Next, we utilize the bond correlation measurements to probe for dynamical heterogeneity, which is characteristic to glass-forming systems~\cite{berthier2011dynamical}. By computing the fluctuations in bond breakages, we obtain a measure of the dynamic susceptibility~\cite{shiba2012relationship, goto2023unraveling}:
\begin{equation}
\chi_b(t) = \frac{1}{N}\sum_{i=1}^{N}\sum_{j=1}^{N} \delta B_i(t)\,\delta B_j(t),
\end{equation}
where $B_i(t)=N_b(0)-N_b(t)$ is the number of broken bonds for the $i$-th ring, $\delta B_i(t)=B_i(t)/2-\langle B(t)\rangle$, $N$ is the total number of rings and the factor $1/2$ avoids double-counting. The average number of broken bonds is
\begin{equation}
\langle B(t)\rangle=\frac{1}{N}\sum_{i=1}^N \frac{B_i(t)}{2}.
\end{equation}

The data for $\chi_b(t)$ are shown in Fig.~\ref{bondcorr}(d),(e). In both systems, $\chi_b(t)$ is non-monotonic, developing a peak whose height grows with density. The increase of the peak amplitude reflects growing dynamical heterogeneity, a hallmark of glass-forming systems~\cite{berthier2011dynamical}. The time at which the peak occurs, $\tau^*$, shifts to longer $t$ with increasing $\rho$, consistent with the increase of $\tau_b$ observed in Fig.~\ref{bondcorr}(c). Comparing the two cases, the stiffness-polydisperse system shows a higher peak, i.e. it exhibits larger dynamical heterogeneity relative to the size-polydisperse system. 

Taken together with the MSD results (Fig.~\ref{MSDplots}), these bond correlation measurements reinforce the conclusion that stiffness polydispersity promotes earlier arrest and more heterogeneous, fragile-like dynamics, whereas size polydispersity delays arrest via geometric frustration and yields a comparatively less fragile glass. We will next analyze the spatial manifestations of the relaxation process. 

\begin{figure*}
	\centering
	\includegraphics[width=2.2\columnwidth]{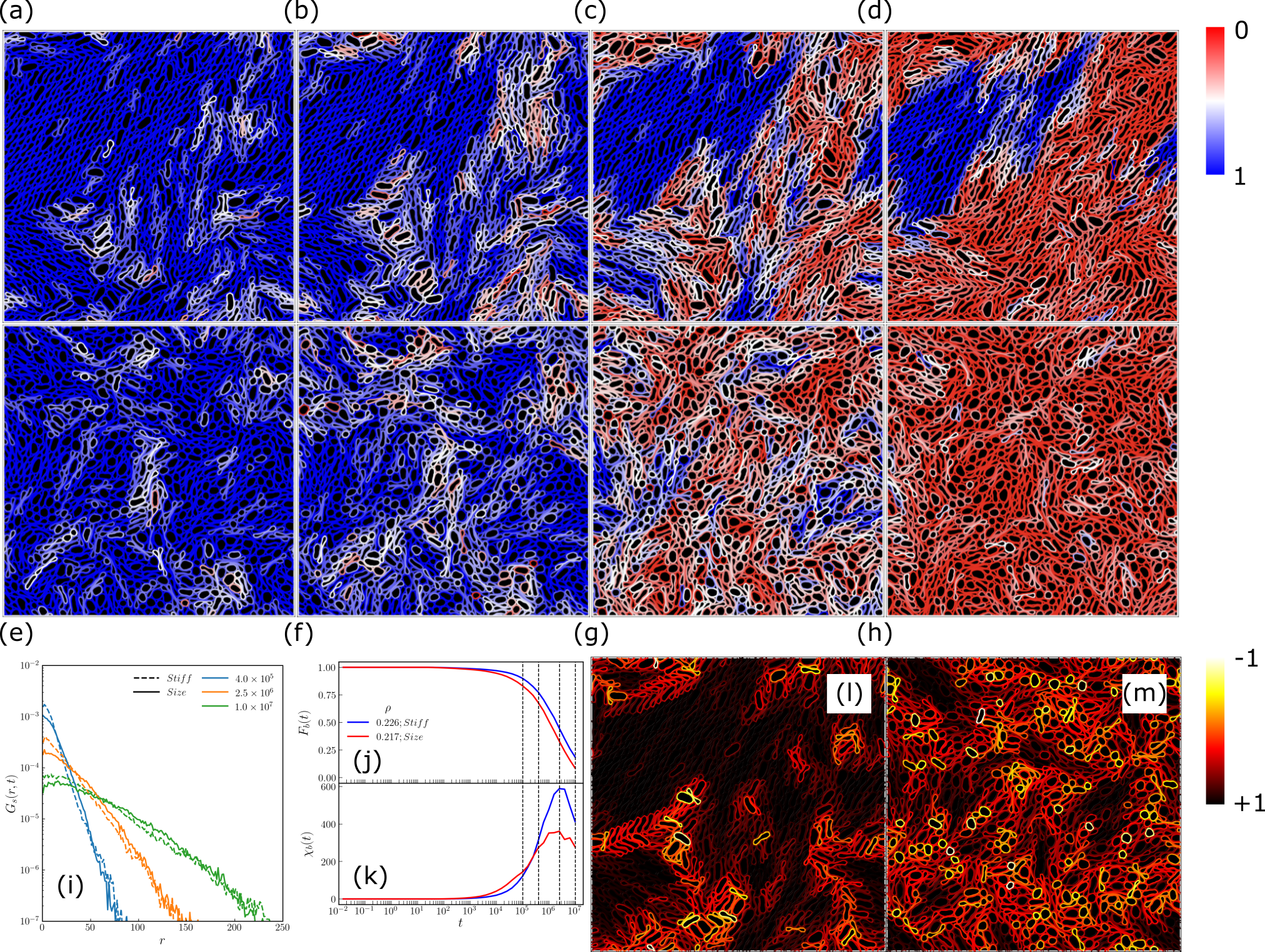}
	\caption{At iso-diffusion point ($D\approx{1.9\times{10^{-4}}}$):  (Top) Spatial maps showing time evolution of local fluidization $X$ (see text for definition) using bond-breakage information, for trajectory corresponding to (a)-(d) stiffness polydisperse  and (e)-(h) size polydisperse system, computed at $t/10^6=0.1, 0.4, 2.5, 10.0$, from left to right, in each case.  The value of local $X$ is shown via the colorbar. (Bottom) (i) Ensemble-sampled self part of van-Hove function, $G_s(r,t)$, for the two systems measured at the time points indicated.   (j) Time evolution of bond correlation function $F_b(t)$ and () its fluctuations $\chi_b(t)$, at the iso-$D$ point.  The vertical dashed lines correspond to the time points at which the maps are computed.  (l)-(m) Configuration at $t=0$, respectively for stiffness and size polydisperse systems, with each ring colored according to its $S_n^\ell$ value.}
	\label{isoD}
\end{figure*}

\subsubsection{Comparing at a iso-$D$ state point}

Utilizing the bond correlation information, we investigate the spatiotemporal features of the dynamics and thereby try to understand the difference in relaxation processes of the two systems. In order to do this comparative analysis, we study the two systems at an iso-diffusive state point, i.e. exhibiting similar approximate values for the diffusion coefficient $D\approx{1.9\times{10^{-4}}}$ (marked in Fig.\ref{MSDplots}(c)). 

To visualize the spatiality of the relaxation process, we construct maps of the ring-resolved bond-breaking correlation function~\cite{scalliet2022thirty}, $F^{i}_b(t)$, where $i$ in the ring index and we color the ring according to the value of $X^i_b(t)=1-F^{i}_b(t)$. Note that $X^i_b(t)$ measures the fraction of virtual bonds that have broken for an individual ring, starting from an initial configuration at $t=0$ when the virtual bonds between neighbouring rings are initialized; thus, $X^i_b(t)$  quantifies the extent of local fluidization. In the top panels of Fig.\ref{isoD}, for the iso-$D$ state points, we show the time evolution of these maps, for both the stiffness (subplots a-d) and size polydisperse (subplots e-h)  systems, at certain time points along the trajectory starting from some equilibrium structure in each case.  This allows us to directly contrast the underlying relaxation mechanisms. 

In Fig.\ref{isoD}(i), we show the corresponding ensemble-sampled self-part of the van Hove function, $G_s(r,t)$, measured at some of the time points for which the maps are shown. The non-Gaussianity of $G_s(r,t)$ evidences the occurrence of heterogeneous dynamics~\cite{chaudhuri2007universal}, signatures of which are visible in the maps shown in Fig.\ref{isoD}(a-h) via the diversity of colors depicting the spatial variation of $X^i_b(t)$ with time. In Fig.\ref{isoD}(j), we show the ensemble-averaged bond correlation function, $F_b(t)$, and also mark the time points at which the maps are generated. Note that although the two systems are nearly iso-diffusive, $F_b(t)$ relaxes slightly slower for the stiffness polydisperse system compared to the size polydisperse system. Further, if one compares the related fluctuations, $\chi_b(t)$, as shown in Fig.\ref{isoD}(k), the stiffness polydisperse systems displays a higher peak, signifying that it is dynamically more heterogeneous. This difference in spatiotemporal heterogeneity is visible in the maps. 

We now focus back on the spatiotemporal signatures of the dynamical fluctuations, for the two cases (Fig.~\ref{isoD}(e)–(h)). At early $t=10^5$, the relaxation of the bonds is  spatially heterogeneous in both cases, with more bonds having broken in the case of the size polydisperse system. The rings for which the bonds have broken form {\it hot spots}. With increasing time, more bonds break and the hot spots proliferate and by $t=4\times{10^5}$, they form pathways of relaxation. By $t=2.5\times{10^6}$, nearly all the rings in the size polydisperse system have lost almost all of their initial neighbours, whereas for the stiffness polydisperse system, a large patch of initial bonds continue to persist. In fact, for the latter system, there are distictly two populations -- for many rings, the bonds have all relaxed, whereas for some they remain intact, which explains why $\chi_b(t)$ shows the largest signature of dynamical heterogeneity around this timescale. Even if we wait till $t=10^7$, we see that some part of this persistent patch still remains, i.e. the dynamics is still heterogeneous, and therefore the corresponding $G_s(r,t)$ continues to be non-Gaussian.

We also try to probe the structural origin of the heterogeneous relaxation. In order to do that, we color rings according to their local orientational order $S_n^\ell$ [Fig.~\ref{isoD}(l),(m)] at $t=0$. Thereafter, if we examine the locations of the early hot spots as revealed by the spatiotemporal maps shown in Fig.\ref{isoD}, we note that in the stiffness-polydisperse system, some of the low-$S_n^\ell$ regions in the initial structure correlate with subsequent fluidization hotspots, demonstrating a clear structure–dynamics link, whereas the dynamical slowing down originates from the more ordered regions ~\cite{kawasaki2007correlation}. On the contrary, such correlations are not evident In the size-polydisperse system.

Overall, Fig.~\ref{isoD} provides a striking visualization of spatial dynamical heterogeneity. Stiffness polydispersity generates persistent coexistence of frozen and fluidized regions, with relaxation predictable from local structure, whereas size polydispersity produces more homogeneous, frustration-driven relaxation that lacks a sharp structural precursor.

\begin{figure*}
	\centering
	\includegraphics[width=2\columnwidth]{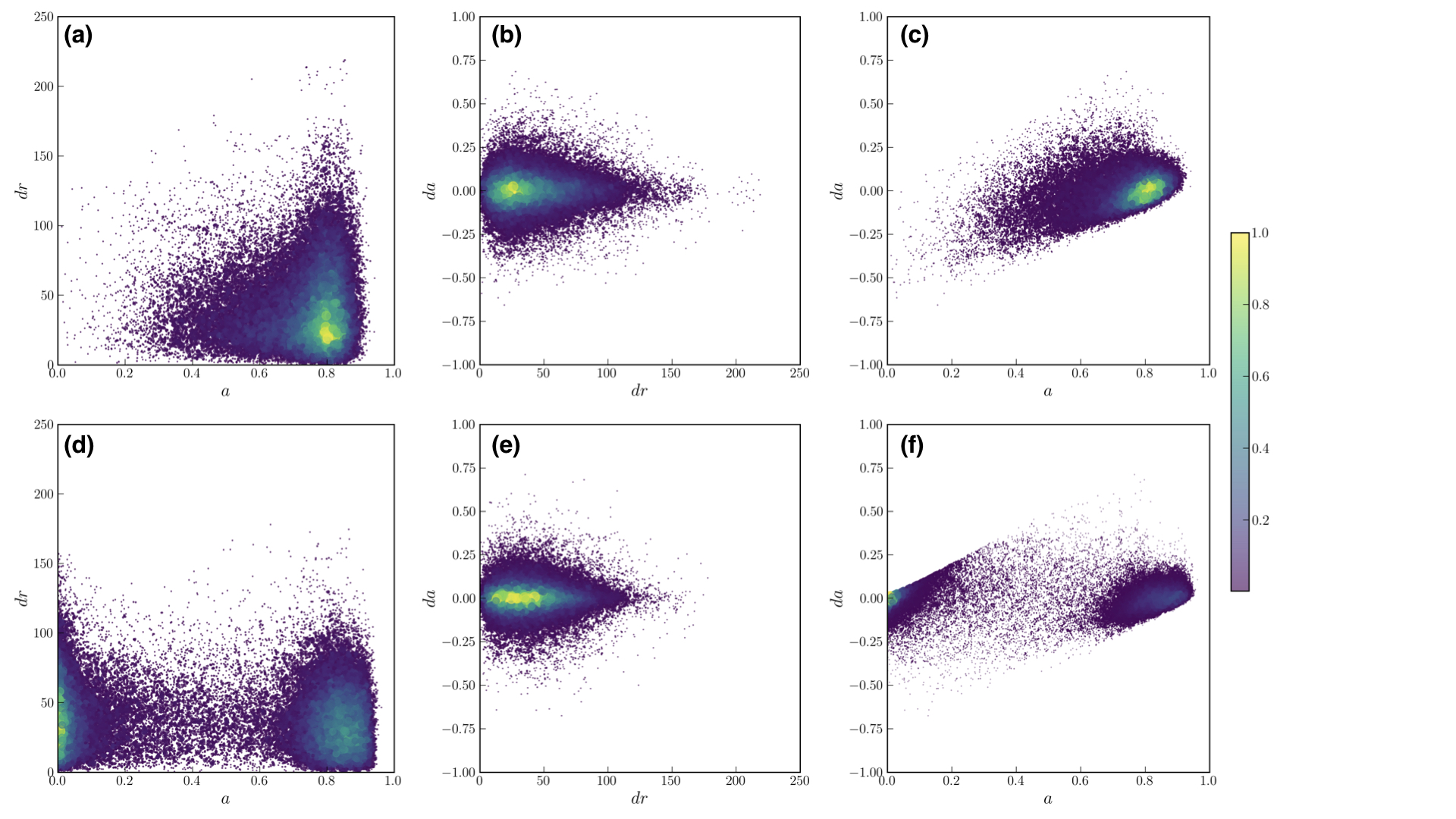}
	\caption{Weighted scatter plot showing correlations between: (a), (d) displacement, $dr$, over timescale $\tau^{*}$, and initial asphericity $a$ at $t=0$; (b),(e) $dr$ and the change in asphericity $da$, over the time window $\tau^{*}$, and (c), (f) the change in asphericity over $\tau^*$ and initial asphericity $a$. Top and bottom panels respectively correspond to stiffnes and size polydisperse systems. The adjoining colorbar provides the scale for the weights visible in the subplots. }
	\label{Scatter}
\end{figure*}

Finally, at the iso-diffusive state point, we examine the interplay of the mobility of the rings with their shapes. In Fig.~\ref{Scatter}, we present weighted scatter plots that link ring displacement $dr$, asphericity $a$, and its change $da$.  Panels (a) and (d) show $dr$, the displacement of the ring over the timescale $\tau^*$ versus the  asphericity $a$ at $t=0$. For stiffness polydispersity, a clear correlation is evident: more aspherical rings undergo larger displacements, reflecting the fact that elongated rings are the primary carriers of relaxation. In contrast, for size polydispersity, both nearly spherical rings (corresponding to small rigid rings) and highly aspherical rings (large deformed rings) contribute to large displacements, while intermediate-sized rings are the least mobile. Panels (b) and (e) examine whether translation $dr$ is related to change in asphericity $da$, both measured over $\tau^*$. In both systems, an anti-correlation is observed: rings that move the least undergo the largest shape changes, whereas the most mobile rings preserve their shape. This indicates that translation and deformation are competing pathways of structural relaxation. Finally, panels (c) and (f) check for the linkage between initial asphericity $a$ at $t=0$ versus the change in shape over $\tau^*$ as captured through $da$. In the stiffness-polydisperse system, less aspherical rings ($a\approx0$) tend to decrease their asphericity, partially recovering a circular shape, whereas more aspherical rings increase their elongation. The same trend is present, though weaker, in the size-polydisperse system. Taken together, Fig.~\ref{Scatter} demonstrates that stiffness polydispersity produces dynamics that are strongly structure-sensitive, with mobility predictable from shape. In contrast, size polydispersity generates mobility from both extremes of the shape distribution, highlighting the role of geometric frustration in driving relaxation.

\section{Discussion and Conclusion}\label{SEC-IV}

Utilizing dense assemblies of semi-flexible ring polymers to model two-dimensional soft glassy systems consisting of deformable objects, we have compared two routes to introducing disorder: stiffness polydispersity (distribution of $K_\theta$ at fixed contour length) and size polydispersity (distribution of contour lengths at common $K_\theta$). At fixed polydispersity ($30\%$) and temperature, increasing density drives slow dynamics in both cases, yet through distinct structural mechanisms. Snapshots (Fig.~\ref{snapshots}) and single-ring shape statistics (Fig.~\ref{asph}) show that stiffness polydispersity produces a unimodal asphericity distribution that shifts to large $a$ at high $\rho$, with most rings becoming rodlike and forming smectic-like bundles (Fig.~\ref{g2_Sn}). In contrast, size polydispersity yields a bimodal $P(a)$ (Fig.~\ref{asph}): small rings remain circular while large rings elongate. This rigidity contrast frustrates ordering (Fig.~\ref{g2_Sn}), consistent with earlier work where deformability promotes dense states \cite{boromand2018jamming, gnan2019microscopic} but dispersity frustrates crystallization \cite{hunter2012physics}. Thus, the size-polydisperse ring polymer assembly can be a promising model for dense complex liquids having a mixture of shapes, sizes, and effective flexibility.

We probe the dynamical behaviour of these two systems using mean squared displacement of the center of mass of the rings (Fig.~\ref{MSDplots}) and bond-breaking correlation functions (Fig.~\ref{bondcorr}). Through these analyses, we demonstrate earlier dynamical arrest and sharper slowdown for the stiffness-polydisperse system compared to the size-polydisperse one. The susceptibility associated with bond correlations develops growing peaks (Fig.~\ref{bondcorr}) with increasing density, evidencing increased dynamical heterogeneity characteristic of glass-forming systems. While prior studies reported both reduced \cite{abraham2008energy} and enhanced fragility \cite{behera2017effects} with polydispersity, our dynamical results in combination with the structural findings clarify that the type of polydispersity matters: stiffness dispersity promotes domain-assisted arrest, whereas size dispersity injects frustration that delays collective arrest.

Our significant analysis is the spatiotemporal comparison of the two systems at an iso–$D$ point in the dense regime (Fig.~\ref{isoD}), which provides a direct real-space demonstration of how the type of polydispersity controls relaxation dynamics in deformable 2D glasses. Maps of local fluidization, using bond-breaking correlations, reveal qualitatively distinct pathways: stiffness polydispersity shows intermittent relaxation with long-lived frozen domains having aligned ordering of the rings coexisting with a highly mobile population, consistent with the larger peak in $\chi_b(t)$; by contrast, size polydispersity quickly generates spatially spanning relaxation tracks with few persistent patches. The structure–dynamics link is further quantified in Fig.~\ref{Scatter}, showing translation and deformation as largely complementary pathways: large $dr$ correlates with small $|da|$, while the sign of $da$ depends on $a$ (circular rings tend to recover, elongated rings further elongate). These trends echo general ideas that dynamic heterogeneity has structural origins \cite{kawasaki2011structural}, but identify specific motifs relevant to deformable rings. In 3D ring systems, arrest arises from threadings and topological constraints \cite{michieletto2016topologically, roy2024bidisperse, roy2022effect}; 2D suppresses these, shifting control to deformability and packing. Our previous work on monodisperse 2D rings showed that flexible rings arrest via crumpling while stiffer rings arrest via local orientational order \cite{ghosh2024onset}. Here we demonstrate that the nature of polydispersity bifurcates these routes: ordered-domain–assisted slowing down for stiffness dispersity versus frustration-delayed arrest for size dispersity. These observations connect to broader soft-particle models where shape and elasticity control jamming and relaxation \cite{boromand2018jamming, gnan2019microscopic}. Future work should explore the rheological response of these systems, which we initiated by shearing binary mixtures of rings \cite{ghosh2025two}. Extension to three-dimensional systems would also reveal how these shape-mediated routes interplay with threading-controlled glassiness \cite{michieletto2016topologically, roy2024bidisperse}.

\section*{Acknowledgment}
 We thank the HPC facility at The Institute of Mathematical Sciences for computing time. PC and SV also acknowledge support via the sub-project on the Modeling of Soft Materials within the IMSc Apex project, funded by Dept. of Atomic Energy, Government of India.

\section*{Data Availability}
The data supporting this article have been included as part of the Supplementary Information. 

\bibliography{2DPoly}

\end{document}